\documentclass{mn2e}
\usepackage{times,epsfig}
\input{psfig.sty}
\begin{document}
         

\def\be{\begin{equation}}
\def\ee{\end{equation}}
\def\mdot{$\dot{m}$ }
\def\kms{\,{\rm {km\, s^{-1}}}}
\def\msun{{$M_{\odot}$~}}
\def\eV{{\rm \ eV}}
\def\ledd{$L_{\rm Edd}$~}
\def\mic{$\mu$ }
\def\se1{s$^{-1}$ }
\def\arcmin{\hbox{$^\prime$}}
\def\arcsec{\hbox{$^{\prime\prime}$}}
\def\degree{$^{\circ}$} 
\def\mic{$\mu$ }
\def\cm2{cm$^2$ }
\def\es{erg~s$^{-1}$ }
\def\gtsima{$\; \buildrel > \over \sim \;$}
\def\ltsima{$\; \buildrel < \over \sim \;$}
\def\prosima{$\; \buildrel \propto \over \sim \;$}
\def\gsim{\lower.5ex\hbox{\gtsima}}
\def\lsim{\lower.5ex\hbox{\ltsima}}
\def\simgt{\lower.5ex\hbox{\gtsima}}
\def\simlt{\lower.5ex\hbox{\ltsima}}
\def\simpr{\lower.5ex\hbox{\prosima}}
\def\la{\lsim}
\def\ga{\gsim}
\def\Lsun{\L_\odot}
\def\cxo{{\it Chandra~}}
\def\nh{N$_{\rm H}$}
\def\lrad{${L}_{\rm r}$}
\def\lx{${L}_{\rm x}$}
\def\ellr{${\ell}_{\rm r}$}
\def\ellx{${\ell}_{\rm x}$}
\def\ie{{i.e.~}}
\def\eg{{\frenchspacing\it e.g. }}
\def\etal{{et al.~}}

\title[BHBs dual radio/X-ray tracks]  
{Assessing luminosity correlations via cluster analysis: Evidence for dual tracks in the radio/X-ray domain of black hole X-ray binaries} 
\author[E. Gallo, B. Miller \& R. Fender]  
{Elena Gallo$^{1}$\thanks{egallo@umich.edu},
Brendan P. Miller$^{1}$
\& Rob Fender$^{2}$\\\\
$^{1}${Department of Astronomy, University of Michigan, 500 Church St., Ann
Arbor, MI 48109, USA}\\
$^{2}${School of Physics and Astronomy, University of Southampton, Highfield SO17 1BJ, UK}}
\maketitle

\begin{abstract}
The radio:X-ray correlation for hard and quiescent state black hole X-ray binaries is critically investigated in this paper. 
New observations of known sources, along with newly discovered ones (since 2003), have resulted in an increasingly large number of outliers lying well outside the scatter about the quoted best-fit relation. Most of these outliers tend to cluster below the best fit line, possibly indicative of two distinct tracks which might reflect different accretion regimes within the hard state.  
Here, we employ and compare state of the art data clustering techniques in order to identify and characterize different data groupings within the radio:X-ray luminosity plane for 18 hard and quiescent state black hole X-ray binaries with nearly simultaneous multi-wavelength coverage. 
Linear regression is then carried out on the clustered data to infer the parameters of a relationship of the form ${\ell}_{\rm r}=\alpha+\beta~{\ell}_{\rm x}$ through a Bayesian approach (where $\ell$~denotes logarithmic luminosities). We conclude that the two cluster model, with independent linear fits, is a significant improvement over fitting all points as a single cluster.  
While the upper track slope ($0.63\pm0.03$) is consistent, within the errors, with the fitted slope for the 2003 relation ($0.7\pm0.1$), the lower track slope ($0.98\pm0.08$) is not consistent with the upper track, nor it is with the widely adopted value of $\simeq 1.4$ for the neutron stars.  The two luminosity tracks do not reflect systematic differences in black hole spins as estimated either from reflection-, or continuum-fitting method. 
Additionally, there is evidence for at least two sources (H1743$-$322 and GRO J1655$-$500) jumping from the lower to the upper track as they fade towards quiescence, further indicating that black hole spin does not play any major role in defining the radio loudness of compact jets from hard black hole X-ray binaries. The results of the clustering and regression analysis are fairly insensitive to the selection of sub-samples, accuracy in the distances, and to the treatment of upper limits. Besides introducing a further level of complexity in understanding the interplay between synchrotron and Comptonised emission from black hole X-ray binaries, the existence of two tracks in the radio:X-ray domain underscores that a high level of caution must be exercised when employing black hole luminosity-luminosity relations for the purpose of estimating a third parameter, such as distance or mass. \end{abstract}

\begin{keywords}
Black hole physics -- Accretion, accretion discs  -- ISM: jets and outflows --
X-rays: binaries -- Radio continuum: general -- Methods: statistical
\end{keywords}

\section{Introduction}
 
The existence of a tight, non-linear correlation between the radio and X-ray flux of hard state black hole X-ray binaries over a wide dynamic range was first reported by Corbel \etal (2003) for GX339$-$4 (see also Hannikainen \etal 1998). 
Re-analyzing a series of quasi-simultaneous observations for a sample of 10 hard state black holes, collected over different epochs and with different instruments, Gallo \etal (2003, hereafter GFP03) concluded that the same relation, with a slope of ($0.7\pm0.1$) {(i)} was a general property of hard state black holes; {(ii)} had similar normalizations for all sources, or in other words, it could be generalized for radio and X-ray {\it luminosities} all the way from quiescence up to a few per cent of the Eddington X-ray luminosity, above which a transition to the thermal dominant state (see, e.g., McClintock \& Remillard 2006 for a review of X-ray states) is typically observed, along with quenching of the core radio emission (Fender \etal 1999; Russell \etal 2011; see Fender 2006 for a review of radio properties of X-ray binaries). 
With the inclusion of the nearby quiescent black hole binaries A0620$-$00 (the faintest radio:X-ray detection on the relation; Gallo \etal 2006) and V404 Cygni at its lowest luminosity level (Corbel \etal 2008), the slope has been more recently revised to a somewhat shallower value ($0.58\pm0.16$). 

\begin{figure*}
\psfig{figure=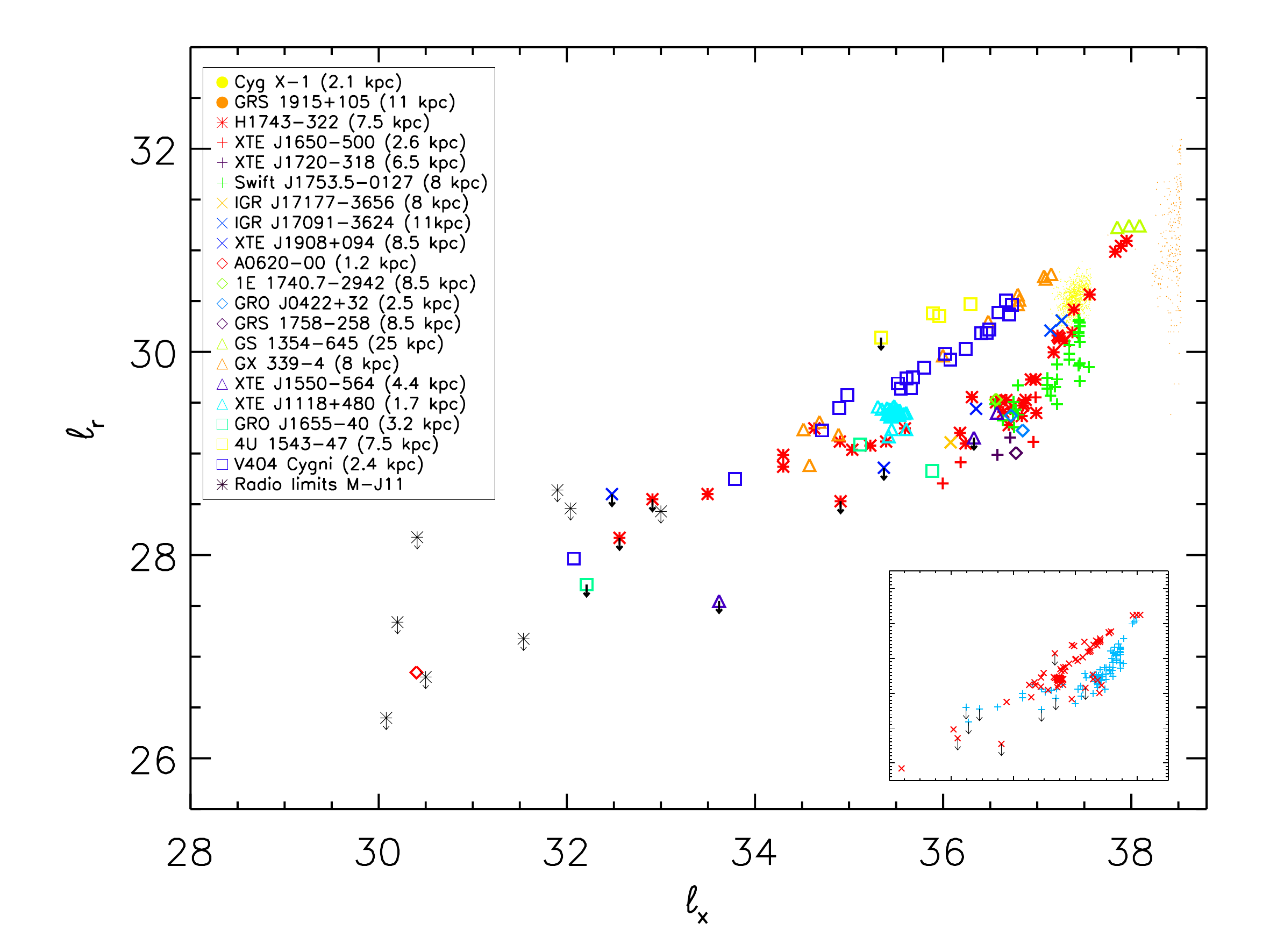,width=0.8\textwidth,angle=0}
\caption{\small The full dataset includes 165 hard-state
observations of 18 different XRBs, with radio and X-ray luminosities
as indicated. Cygnus X-1, GRS 1915+105, along with the {radio} upper limits (combined with strictly non-simultaneous X-ray detections)  by Miller-Jones \etal (2011) are plotted for
illustrative purposes but are not included in our analysis (see text).  $\ell$
denotes $log$ luminosities in units of \es. For the 18 sources under analysis, the inset illustrates 
luminosities for the sources with secure distance measurements, in red, vs. uncertain distance measurements, in blue.}\label{fig:all}
\end{figure*}

As radio and X-ray luminosities are thought to trace different emission processes -- namely self-absorbed synchrotron from downstream in the jet vs. inverse Compton/optically thin synchrotron from the compact corona/jet base, respectively --  the slope of the correlation has been interpreted as evidence that a dominant fraction of the X-ray spectrum of hard state black hole X-ray binaries is due to optically thin synchrotron emission from the jet base (Markoff \etal 2003). Broad-band (radio-to-X-ray) spectral fitting provides independent evidence that this might be the case for a variety of weakly accreting black holes, both stellar and super-massive (Markoff \etal 2001, 2004, 2005, 2008; Maitra \etal 2009, 2011; Migliari \etal 2007; Russell \etal 2010; Malzac \etal 2004 -- see e.g. Corbel \etal 2004 for a counter-example, however), although other possible interpretations of the scaling relation remain open (e.g. Maccarone 2005).    
Regardless of its physical explanation, the non-linearity of such `universal' scaling implies the existence of an accretion regime where most of the accretion power is released (likely in the form of mechanical power) within the jet, as opposed to being dissipated locally within the accretion-corona system (as discussed in GFP03, this is simply as a result of the empirical 0.7 slope combined with the theoretical prediction that the radio power and the total jet power correlate with a $1.4$ power law slope in partially self-absorbed jet models; cf. Blandford \& K\"onigl 1979). 

Migliari \& Fender (2006) subsequently reported on a similar investigation for the neutron star systems, showing that a steeper (albeit less tight) correlation exists between the radio and X-ray luminosity in neutron stars, too, of the form $L_{\rm r}\propto L_{\rm X}^{~1.4}$. Incidentally, the difference in the inferred slopes for the black holes and the neutron stars nicely combine such that, unlike the black holes, neutron stars never enter the `jet-dominated' regime. If so, that could be sufficient to account for the observed gap in luminosity between quiescent black holes and neutron stars, without requiring the presence/absence of a solid surface (Fender \etal 2003; see also K\"ording \etal 2006a). \\
A qualitatively similar relation, with a slope of ($0.6\pm 0.1$), has been reported between the optical-infrared (IR) and the X-ray luminosity of both black hole X-ray binaries and low mass neutron stars X-ray binaries while in the hard state (Homan \etal 2005, Russell \etal 2006, Coriat \etal 2009), indicating a non-negligible contribution from the jet to the near-infrared flux in these systems, although the relative contributions from jet vs. reprocessed disc emission vary greatly between the two types of sources. Interestingly, there appears to be a hysteresis effect in the near-IR:X-ray correlation of hard state black hole X-ray binaries, namely, at a given X-ray luminosity, the near-IR emission appears to be weaker during the rise of an outburst than during the decline of an outburst (Russell \etal 2007). 

Expanding on those works, two separate groups examined the possibility that similar scalings hold for super-massive black holes in Active Galactic Nuclei (AGN) of the radio-weak variety, i.e. the AGN sub-class whose properties most closely resemble those of hard state black hole X-ray binaries (e.g. Maccarone \etal 2003, K\"ording \etal 2006b). Merloni \etal (2003) and Falcke \etal (2004), independently came to the same conclusion, that is, that weakly accreting black holes -- from stellar mass scales to super-massive -- preferentially lie on a plane in the three-dimensional parameter space defined by their mass, nuclear X-ray luminosity and radio luminosity (see K\"ording \etal 2006c, Merloni \etal 2006, Li \etal 2008, G\"ultekin \etal 2009, Yuan \etal 2009, Miller \etal 2010 for more recent studies). While the physical explanation for this finding is non trivial (see e.g. Heinz 2004, Heinz \& Merloni 2004, Plotkin \etal 2012), the best-fit relation for this `fundamental plane of black hole activity' may be employed for estimating or predicting one of the three observables (most notably mass) based on measurements of the remaining two (see e.g. Reines \etal 2011 and Miller \& G\"ultekin 2011 for two recent as well as remarkable examples). \\
Going back to the stellar mass black holes in X-ray binaries, measurements of radio and X-ray flux for newly discovered sources are similarly employed to estimate their most likely distance based on their location with respect to the best-fit relation in the radio/X-ray domain (e.g. Cadolle-Bell \etal 2007, Paizis \etal 2011, Rodriguez \etal 2011).
Despite its fairly broad impact, though, the initial picture of a tight universal radio:X-ray luminosity relation has slowly drifted toward a more complex scenario. 
An increasing number of `outliers' (Gallo 2007) have been reported, defined as falling well outside the scatter about the original GFP03 best-fit relation (see next section for references). Most of these points tend to be clustered {\it below} the best-fit relation (i.e. at lower radio luminosities for a given X-ray luminosity), leading to the suggestion that there may be two separate tracks in the radio/X-ray luminosity plane of black hole X-ray binaries, each representing a different mode of accretion within the hard state (e.g. Coriat \etal 2011; see also Rushton \etal 2010, restricted to GRS~1915$+$105). 

Motivated by Occam's razor, this paper sets out to give a statistically rigorous answer to the question whether, in the face of an increasing number of outliers and uncomfortably large scatter, it still makes sense to refer to a universal radio/X-ray correlation for hard state black holes X-ray binaries (hereafter, BHBs). The reader is referred to Xue \& Cui (2007) for a different approach.

\section{Sample description}
We consider a sample comprised of multiple, {\it quasi-simultaneous} (i.e. within a day) radio
and X-ray observations of 18 black hole X-ray binary systems while in
the hard X-ray state, for a total of 166 data points (157 detections and 9 upper limits in radio luminosity).  In addition to the original sample of hard
state BHBs discussed in GFP03 (9 sources excluding Cygnus X-1\footnote{Data points relative to the high mass BHB Cygnus X-1 are
excluded from this analysis (as they were from GFP03) on the basis of its repeated `failed' state transitions. See e.g. Rushton \etal (2012), discussing the peculiar radio behavior of this source.}), the following 9 are included: XTE J1650$-500$ (data from Corbel \etal 2004); XTE J1908+094 (data from Jonker \etal 2004); A0620$-$00 (data from Gallo \etal 2005); XTE~J1720$-$318 (data from Brocksopp \etal 2005, 2010); GRO~J1655$-$40 (data from Migliari \etal 2007 \& Calvelo \etal 2010); IGR~J17177$-$3656 (data from Paizis \etal 2011); Swift J1753.5$-$0127 (data from Cadolle-Bel \etal 2007 \& Soleri \etal 2010); H1743$-$322 (data from McClintcok \etal 2009, Jonker \etal 2010, Coriat \etal 2011); IGR~J17091$-$3624 (data from Rodriguez \etal 2011). 
In addition, new data points are included for 4U~1543$-$47 (data from Kalemci \etal 2005) and V404 Cygni (data from Corbel \etal 2008).  
A plot of radio luminosities (calculated by integrating up to the observed frequency and assuming a
flat spectrum) versus 1-10 keV X-ray luminosities\footnote{For
consistency with GFP03, unabsorbed luminosities quoted in different
energy bands have been converted to 1-10 keV luminosities using
WebPimms and assuming a power law spectrum with index $\Gamma=2$.} for
the 18 hard state BHBs under
consideration is shown in Figure~\ref{fig:all}, whose label also includes the adopted distances. Data points from Cygnus X-1 and GRS~1915+105 are also included for comparison, along with the radio upper limits from a recent deep survey of quiescent BHBs carried out by Miller-Jones \etal (2011, M-J11 hereafter), but for which simultaneous X-ray counterparts are based on archival X-ray detections, often from observations that were taken several years prior to the radio survey.   

Several authors (e.g. Rushton \etal 2010, Jonker \etal 2010, Coriat \etal 2011, Ratti et al., submitted) have suggested
that there appears to be dual upper and lower luminosity tracks in this domain,
particularly when compared against figure 2 of GFP03, where a single
correlation was identified. For descriptive purposes, the divide can be loosely set by the ${\ell}_{\rm r}={\ell}_{\rm x}-7$ line, where $\ell$
denotes $log$ luminosities in units of \es. We investigate the effective validity of such a split
classification in the following sections. 

When available, luminosities have been updated
based on recent and or more accurate distance measurements (based on the references in 
Table~2 of Fender \etal 2010 and Jonker \& Nelemans 2004, except for the distance to XTE~J1550-564 which has been recently revised to 4.4 kpc; cf. Orosz \etal 2011).  Seven of out the 18 sources considered here (i.e. H1743$-$322, XTE J1650$-$500, XTE J1720$-$318, XTE J1908+094, Swift J1753.5$-$0127, IGR J17177$-$3656 and IGR J17091$-$3624) have less than reliable distance values, in that their distance determinations are based on loose absorption arguments, or, in some instances, on their very location on the radio/X-ray luminosity plane investigated here. We refer to this subsample as `uncertain distance' sample (vs. `secure distance sample'). The inset of Figure~1 illustrates the locations of the two samples: note that {culling objects with uncertain distance measurements removes the majority of the lower
track data points (71 out of 77), although those six points that remain come from 3 distinct
objects with secure distances} (GRO~1655$-$40,  XTE~1550$-$564 and GRS~1758$-$258). 


\section{Clustering analysis}

For the sample described above, we address three general questions: (i) whether there are necessarily
two (or more) clusters populated within the ${\ell}_{\rm
r}-{\ell}_{\rm x}$ plane; (ii) if so, what function
optimally characterizes the clustering, and with which cluster is each
individual observation most likely associated; (iii) given these
groupings, is the single linear regression model derived from all
observations consistent with the best-fit lines within each cluster,
and to what degree, if any, is the distinct-trends model statistically
preferred. We wish to stress that the clustering analysis can only handle {\it detections}; we include the treatment of the upper limits in the linear regression analysis. 

\begin{figure}
\psfig{figure=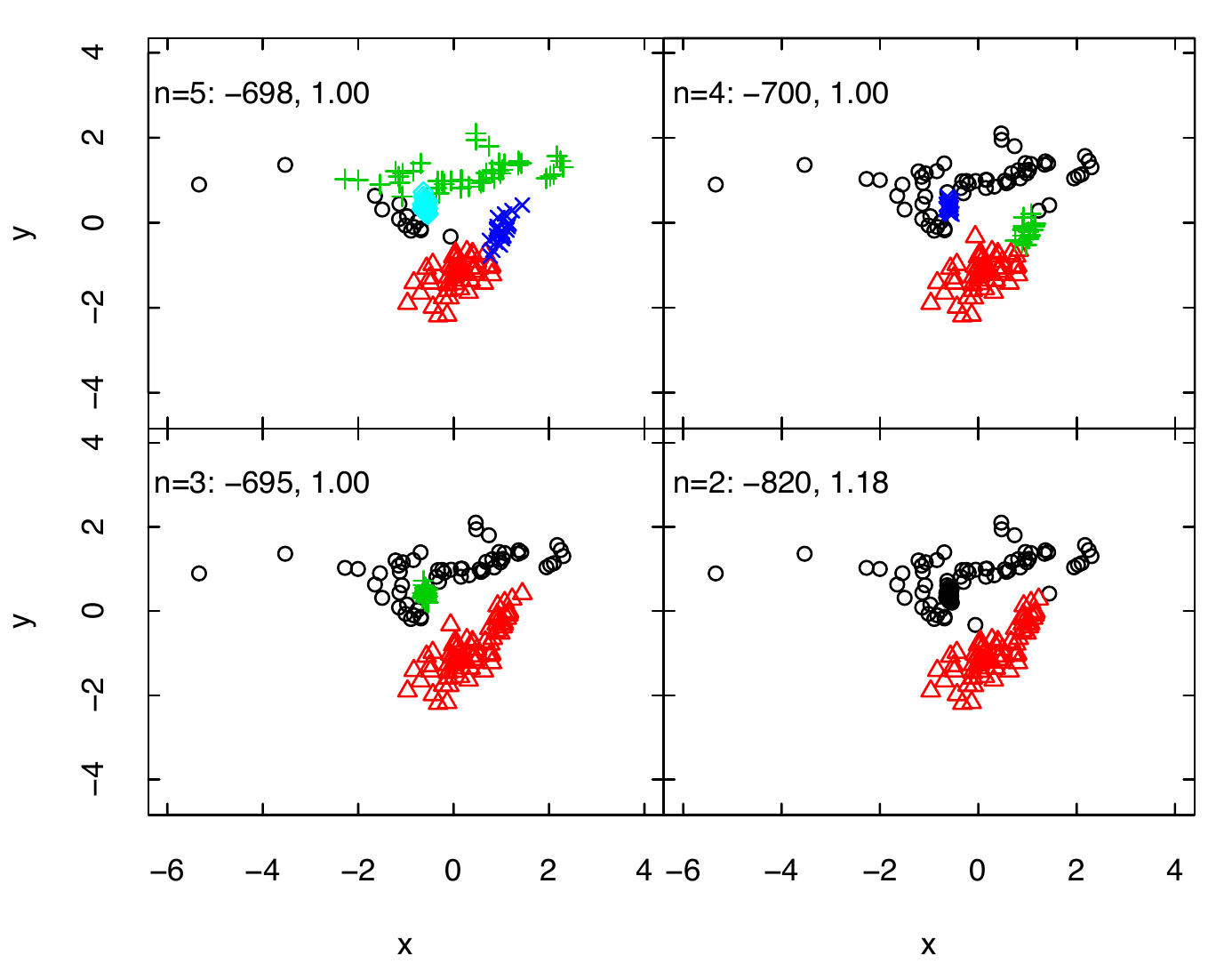,width=0.45\textwidth,angle=0}
\caption{\small Model-based clustering results for $n$=2,3,4,5
clusters ({\it mclust\/} code). Distributions are typically ellipsoidal, with varying
enclosed area and orientations. Labels give the Bayesian Information
Criterion (BIC) ans sclaled-BIC values; models with the lowest (i.e. =1) scaled-BIC are preferred.}\label{fig:mclust}
\end{figure}

\begin{figure}
\psfig{figure=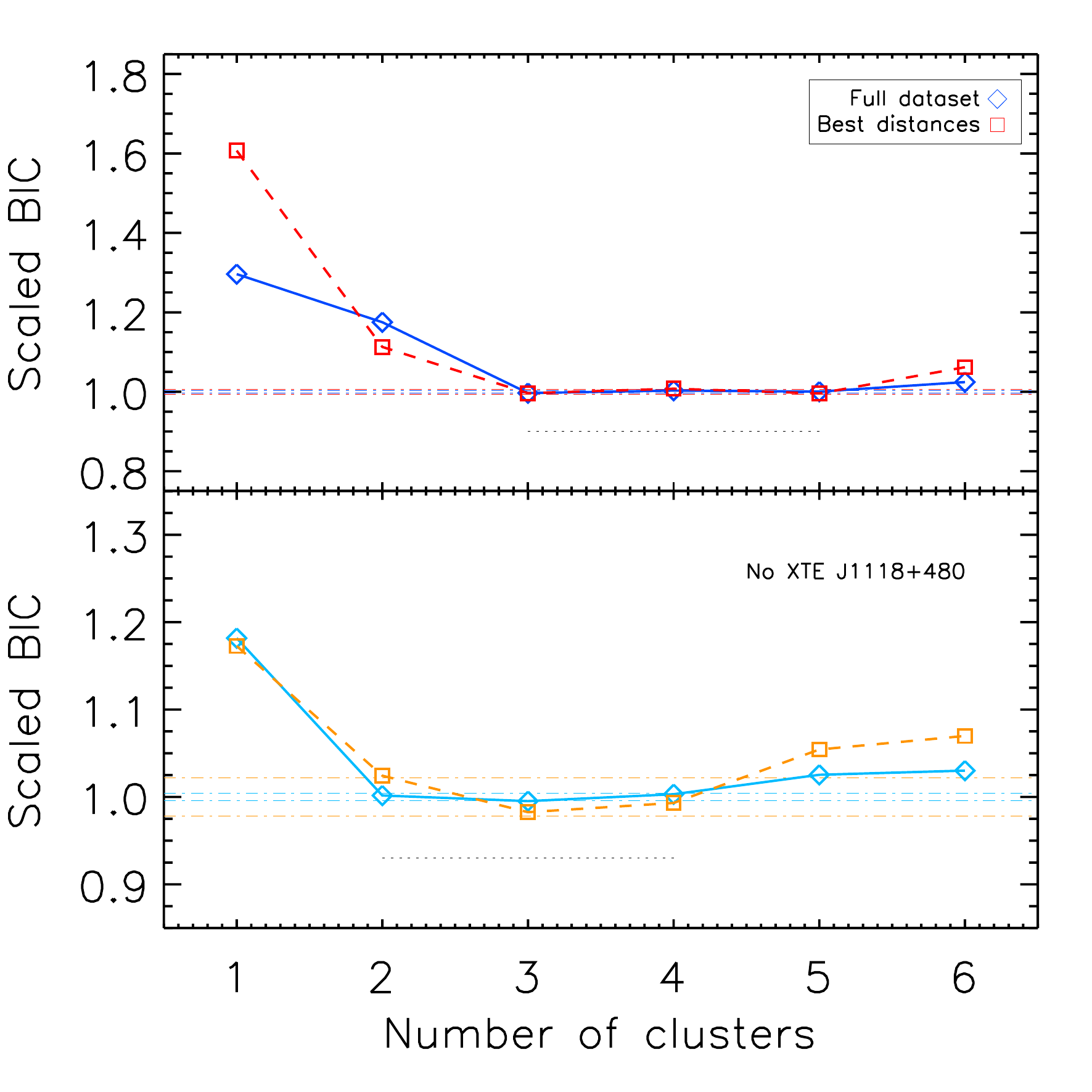,width=0.45\textwidth,angle=0}
\caption{Scaled BIC values (i.e. divided by the BIC value corresponding to the best-fit cluster model) for $n=1, 2, 3,..6$ model-based clusters; dashed and dot-dashed lines
are 1$\pm\sigma$ over the interval marked by the dotted segment.  A measure of the preferred number of clusters is given by {the leading edge of the plateau in scaled BIC}.  
With (without) XTE J1118+480, three (two) clusters
are preferred.}\label{fig:bic}
\end{figure} 


Potential clustering of observations within the radio/\hbox{X-ray} luminosity plane is investigated using standardized,  {\it $x$:$y$} coordinates (as opposed to raw \lrad:\lx).  
`Standardization' is a way of scaling variables so that different variables, measured in different units and/or over different dynamic range, can be more easily compared. While both radio and X-ray luminosity for our sample are measured in \es, they are integrated over different energy ranges and likely account for different fractions of the bolometric luminosity emitted via their underlying emission process; in addition, the data set tends to stretch along the \lx~axis. 
The first transformation ($L_{\rm X} \rightarrow x'$, $L_{\rm r} \rightarrow y'$) includes taking the logarithm, subtracting the mean and dividing by the standard deviation, such that the resulting range of values for the primed coordinates is comparable along the $x'$ and $y'$ axis, while they both end up centered at $x'=y'=0$. 
Principal Component Analysis (PCA; e.g. Jolliffe \etal 2002) applied to $x'$ and $y'$ identifies the major component vector (essentially parallel to the upper track as defined above) to account for 87\% of the data variability. The data are rotated accordingly around their mean, such that they end up aligning with the principal component. The rotated variables are again scaled to unit variance, and clustering
analysis is finally carried out on the resulting vectors, which we will refer
to as $x$ and $y$ hereafter (see Figure~\ref{fig:mclust}).  This ensures comparable sensitivity (in terms of the ability to identify clusters) both along and across luminosity
trends. \\

\subsection{Model-based clustering}

We first use {\it mclust\/}, a multivariate normal mixture modeling and 
model-based clustering code (Fraley \& Raftery 2002, 2007), to evaluate the presence of $n$ clusters, where $n$ is allowed to vary 
from one to six. The considered models are spherical,
diagonal, or ellipsoidal distributions (permitted to vary in area,
shape, and orientation, where applicable), and models are ordered by
likelihood as estimated using the Bayesian Information Criterion
(BIC).  BIC resolves the problem of `over-fitting' (increasing the likelihood of a model by adding parameters) by introducing a penalty term for the number of parameters in the model (it can be thought as an analog to reduced $\chi^2$ statistics in a clustering analysis framework). Given a series of models, the one with the higher BIC value is the one to be preferred (note that, by construction, BIC$<$0).

Figure 2 illustrates the results of this exercise by showing different clustering for $n=2,3,4,5$ (bottom right, bottom left, top right and top left, respectively). At face value, three clusters, with ellipsoidal distributions, are preferred by the data (with a best-fit BIC value of $-$695). For simplicity, the estimated BIC values can be scaled to the best-fit BIC, such that a scaled-BIC of unity corresponds to the best model.  
Figure~\ref{fig:bic} illustrates the variation of such scaled BIC value as a function of number of clusters for the full/secure distance samples. In this representation, a measure of the preferred number of clusters is given by {the leading edge of the plateau in scaled BIC}.  
While the two cluster model (essentially identical to the upper and lower tracks as defined in Section 1) is associated with a slightly higher (1.18 vs. 1) scaled-BIC compared to the three clusters, it is clear that {\it two clusters provide a better description of the data set over a single cluster model} (BIC value 1.18 vs. 1.3, respectively).
In addition, the three clusters consist of the upper and lower tracks as well as a separate grouping 
containing XTE J1118+480 points, which is likely distinguished by the
particularly dense sampling over a very narrow dynamic range for this object, rather than any physical
characteristics (for fitting purposes, the same XTE J1118+480 data points discussed here were actually averaged to a single data point in GFP03). 

While four and five clusters (splitting the upper and then
lower track) do not provide an improvement over three, when restricting consideration to the subset of objects with secure
distance measurements, the preferred three cluster model again
isolates most of the XTE J1118+480 points but now splits off the
high-luminosity segment of the upper track while merging the
low-luminosity segment of the upper track with the (here sparsely
populated) lower track.

\begin{figure}
\psfig{figure=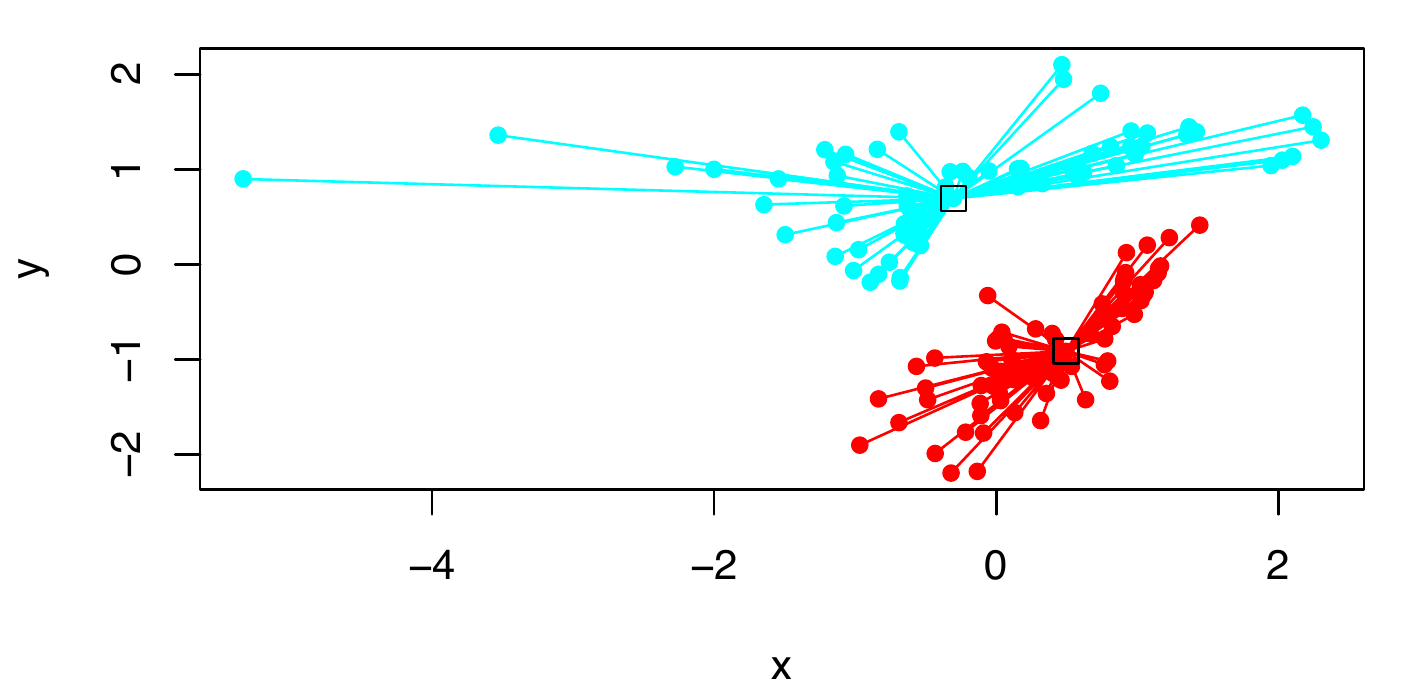,width=0.46\textwidth,angle=0}
\psfig{figure=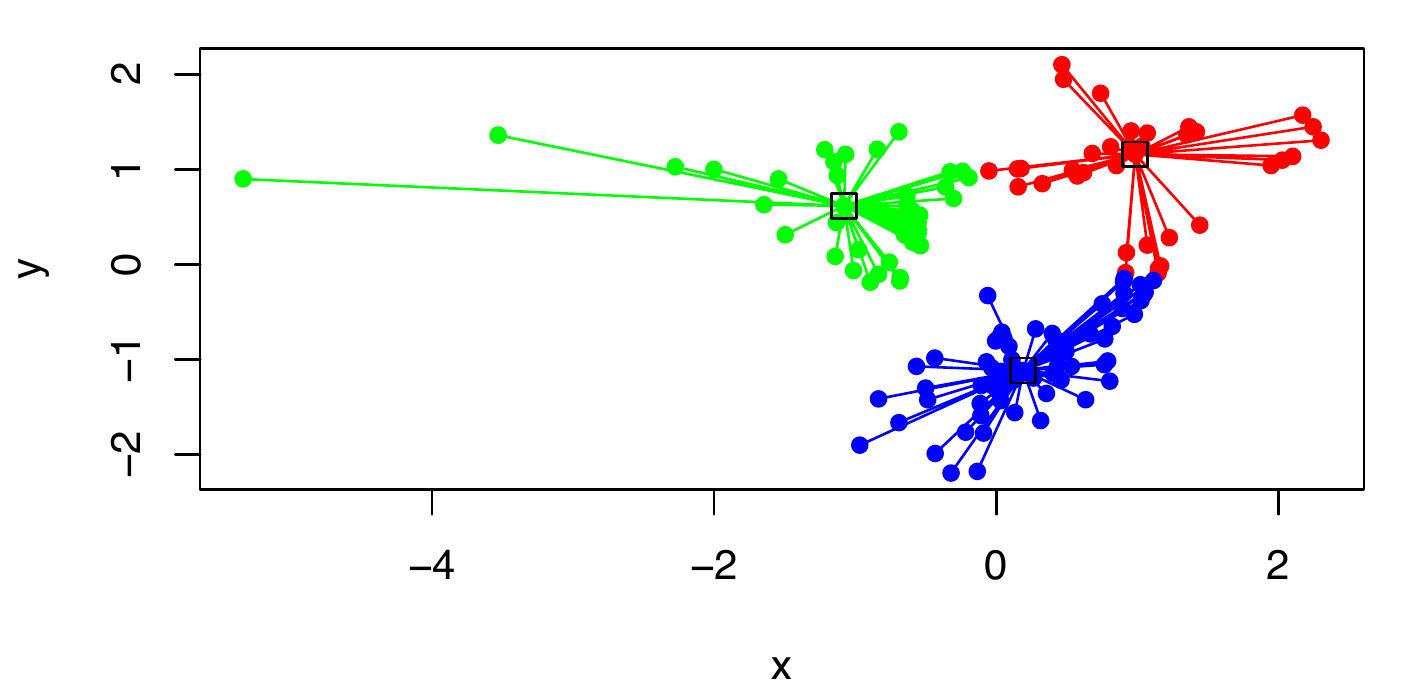,width=0.46\textwidth,angle=0}
\caption{Clustering derived from affinity propagation ({\it apcluster}) with
initial preferences $p$=$-$100 and $p$=$-$70, respectively, for the top and bottom panel. The damping factor $\lambda$ is set to 0.9 for both cases. The exemplars (i.e. cluster centers as identified by the code) are marked
with squares.\label{fig:aff}}
\end{figure}

In summary, this method identifies three clusters -- one corresponding entirely to the low-dynamic range/high cadence XTE J1118+480 data -- as preferred for
the full dataset and likewise for the subset of objects with secure
distances, although in the latter case the groupings are not obviously
related to the upper and lower tracks. When XTE J1118+480 is averaged to a single data point,
two clusters (the upper and lower tracks) are preferred for the full
dataset, and perhaps also for the best distances subset (splitting off
the high-luminosity segment of the upper track), although here three
clusters (now separating out the lower track) does still have a slightly better
BIC value.\\

An independent method of assessing clustering significance (also
without the necessity of fixing {\it a priori} the clustering to be
tested) is provided by the heirarchical Bayesian model implemented by
Fuentes \& Casella (2009; {\it bayesclust\/}). 
{\it Bayesclust} allows to first test the null hypothesis (equivalent to just one cluster). It then procedes to test the alternative hypothesis to be either 2, 3 or 4 clusters at any one time. 
The assumed
distribution for the observations is multivariate normal, and we use the default non-informative priors on $\sigma$ (the variance) and $\mu$ (the mean for each cluster), with a minimum cluster size of 10\% of objects
(i.e., 16 data points) enforced. 
{\it Bayesclust} adopts an importance sampling estimate for the Bayes factor by sampling from the partition distribution $g(\omega)$. 
The significance of clustering is then 
assessed through a frequentist approach, i.e. by simulating the distribution of the Bayes factors under the null hypothesis of a single cluster.
The corresponding $p$-value, which gives a measure of the likelihood for 1 cluster, is 
is 0.003, indicating again that two clusters are preferred over one. The suggested partitioning, however, is somewhat different from what identified by {\it mclust}, in that a small fraction of the data points on the high luminosity portion of the bottom track as identified by {\it mclust} are identified as members of the top track by {\it bayesclust}. 

\subsection{Optimal clustering of observations}

We further test and verify the model-based clustering presented above through applying three additional approaches:
partitioning around medoids (Kaufman and Rousseeuw 1990; {\it pam}), affinity propagation through
iterative maximization of net similarity (Frey \& Dueck 2007;
implemented by Bodenhofer \etal 2011 as {\it apcluster}), and
hybrid hierarchival clustering using mutual clusters (Chipman \&
Tibshirani 2006; {\it hybridHclust}). 

\begin{figure}
\epsfig{figure=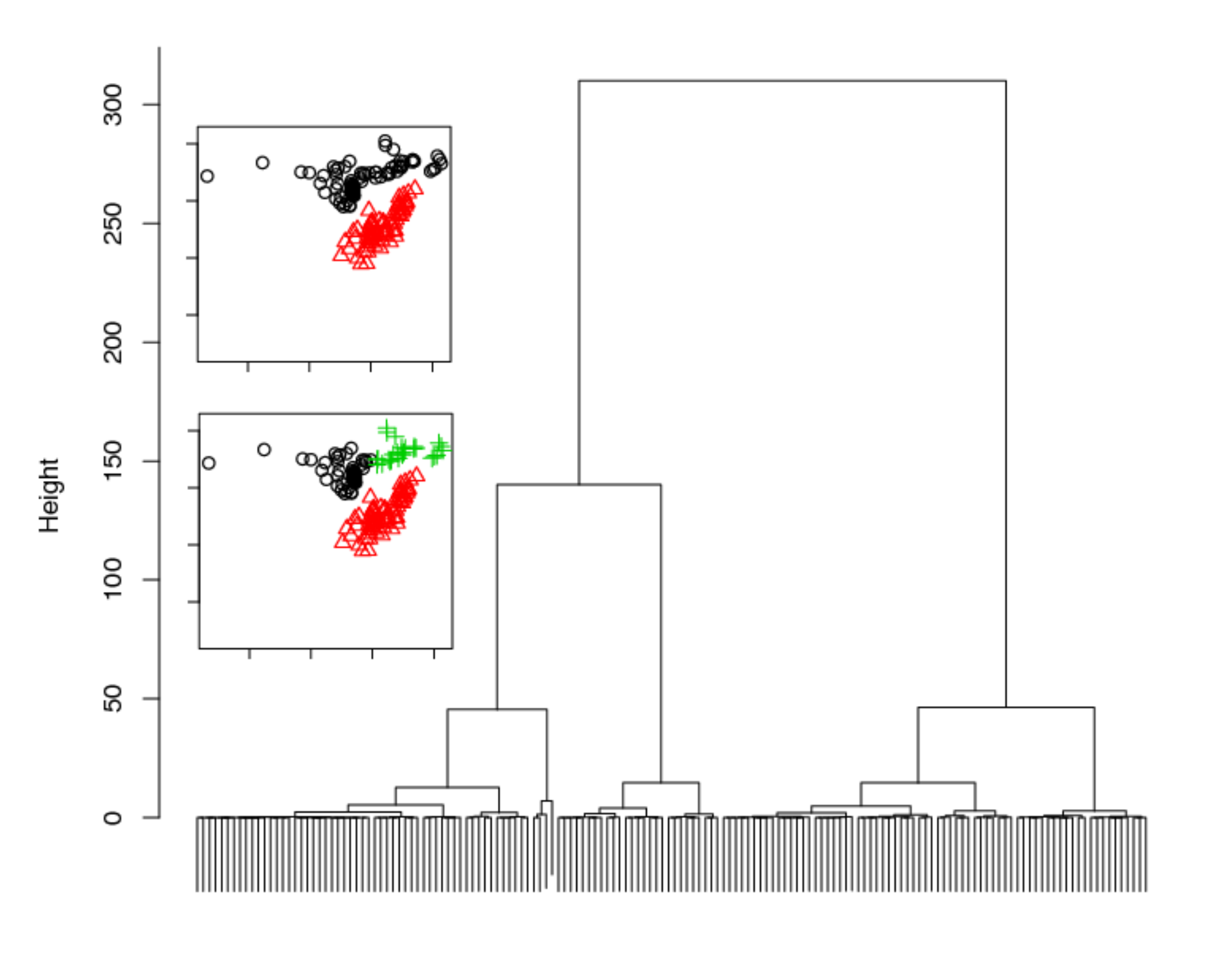,width=0.52\textwidth,angle=0}
\caption{Dendrogram constructed through top-down hybrid hierarchical clustering,
preserving mutual clusters ({\it hybrydHclust}). The inset shows the corresponding two and three cluster splits (which have been adopted for the linear regression analysis).}\label{fig:dendo}
\end{figure} 

Both the k-means and k-medoids algorithms are partitional, i.e. break the dataset up into groups, and both minimize the distance between points labeled to be in a cluster and a point designated as the center of that cluster, although partioning around medoids is considered a more robust version of the standard
$k$-means clustering algorithm in that it is more robust to noise and outliers. {\it Pam} is a classical partitioning technique that clusters the dataset of $n$ objects into $k$ clusters, where $k$ is a user-supplied parameter.  
When two clusters are requested, the code again separates out the upper and lower tracks\footnote{Selecting from
${\ell}_{\rm x}, {\ell}_{\rm r}$ rather than from $x, y$ forms instead
two groups split near ${\ell}_{\rm x}=36.3$.}. As before, the most
luminous points are included with the upper track, although they
appear potentially also consistent with the lower track. When three
clusters are requested, the upper track splits, but the
high-luminosity end of the lower track also joins the high-luminosity
segment of the upper track.

{\it Apcluster} (Frey \& Dueck 2007) is a fairly new algorithm that takes as input measures of `similarity' between pairs of data points and simultaneously considers all data points as potential cluster centers. In complex multi-dimensional problems, such as grouping images of faces, of genes identification, affinity propagation significantly out-performs even the best
result of multiple $k$-means runs. 
Cluster centers are usually found by randomly choosing an initial subset of data points and then iteratively refining it, making the final result somewhat dependent on whether the initial choice is a fair representation of the whole data set. Instead, affinity propagation identifies exemplars (i.e. cluster centers) via `message passing' between points\footnote{Two kinds of message are exchanged: {(i)} the `responsibility' $r(i,k)$ (sent from data point $i$ to candidate exemplar point $k$) reflects the accumulated evidence for how well-suited point $k$ is to serve as the exemplar for point $i$, and {(ii)} the `availability' $a(i, k)$ (sent from $k$ to $i$), which reflects the evidence for how appropriate it would be for point $i$ to choose point $k$ as its exemplar, taking into account the support from other points that point $k$ should be an exemplar (from Frey \& Dueck 2002).}. 
Messages are exchanged between data points until a high-quality set of exemplars and corresponding clusters emerge. Within the algorithm framework, adding a tiny amount of noise to the similarities to prevent degenerate situations (oscillations) corresponds to increasing the damping factor $\lambda$ (allowed to vary between 0 and 1). High values of the preference $p$ will cause affinity propagation to find many exemplars, while low values will lead to a small number of exemplars. 
Going back to our data set,  low initial preference ($p$=$-$100) identifies again the upper and lower tracks; with
higher preference ($p$=$-$70), the split occurs midway along and transverse to the
upper track. Changing the damping factor from $\lambda$=$0.9$ to $\lambda$=$0.5$ results in a slightly different membership for the 3 cluster model. 

Finally, we test a hybrid hierarchical clustering method that uses mutual clusters, where a mutual cluster is composed of points having a maximum relative
distance less than the shortest distance to any point outside the
group. {\it HybridHclust} relies on the fact that bottom-up clustering cannot break a mutual cluster. That is, when agglomerating points, these method will never add points outside the mutual cluster before first joining all points inside the mutual cluster.
The resultant top-down clusterings are then stitched together to form a single top-down clustering diagram, known as `dendrogram', which allows to identify the number and composition of the clusters. 
The dendrogram for our data set is shown in Figure~\ref{fig:dendo}, with insets illustrating the two and three
cluster branches. The two primary clusters (splitting at a height of
300) are identified with the upper and lower tracks. The next two
sub-clusters (splitting at a height of 140) dissociate the upper track
into higher and lower luminosity portions. \\

Synthesizing the results from the two previous Sections, {\it we conclude that two physically meaningful
clusters are indeed significantly present in the radio/X-ray domain of hard state BHBs}, loosely corresponding to the upper
and lower tracks as identified in Section 2, although the specific characteristics of the lower
track in the full dataset are influenced by objects with
uncertain distances.  It is worth noticing that the six most luminous points in the plot (corresponding to \ellr$>$31~{and}~\ellx$>$37 in Figure~\ref{fig:all}) are identified as members of the upper track by all clustering algorithms, whereas they could easily be classified as the upper luminosity end of the lower track in a more arbitrary `by-eye' classification.    
In order to assess the robustness of our clustering analysis on measurement errors -- which are not included in any of the routines discussed above -- we verified that randomly scrambling the data points within a Gaussian of $\sigma=0.15$ in ${\ell}_{\rm x}$ and ${\ell}_{\rm r}$ (i.e., a factor of
0.7--1.4 in flux, which may account for intrinsic source variability, lack of strict simultaneity, as
well as measurement errors) still returns two clusters as the preferred model, and that the cluster membership is preserved. 

We adopt the results described above to populate the cluster members for the linear regression analysis described
in the following Section. The two cluster model -- as identified by all methods except for {\it bayesclust} -- corresponds to the lower and upper tracks in, e.g., the top inset of Figure~\ref{fig:dendo}, while the three cluster model typically subdivides the upper track, although it should be kept in mind that various methods suggest somewhat different three-cluster splits ({\it apcluster} and {\it hybridHclust} return exactly the same groupings for the three cluster model, as shown in the bottom panel of Figure~\ref{fig:aff} and the bottom inset of Figure~\ref{fig:dendo}, respectively).  
For the two cluster model, the various algorithms ({\it mclust, apcluster, pam} and {\it hybriHclust}) assign 86 and 71 data points to the upper and lower clusters, respectively. Restricting the selection to the secure distance sub-samples returns 74 and 6 points.

\section{Linear regression within clusters}

We procede to carry out linear regression on $({\ell}_{\rm r}-29.5)=\alpha+\beta({\ell}_{\rm x}-36.6)$, with intrinsic random
scatter included. 
The centering is based on the median luminosities in
the full sample. Two Bayesian modeling packages are used to determine
the parameters, the first from Kelly (2007; implemented in IDL as
{\it linmix\_err.pro\/}), and the second from Hall (2011; implemented
in $R$ as {\it LaplacesDemon\/}\footnote{http://www.statisticat.com/ laplacesdemon.html}), which serves for comparative purposes with analysis performed on the detections only.   

\begin{figure}
\psfig{figure=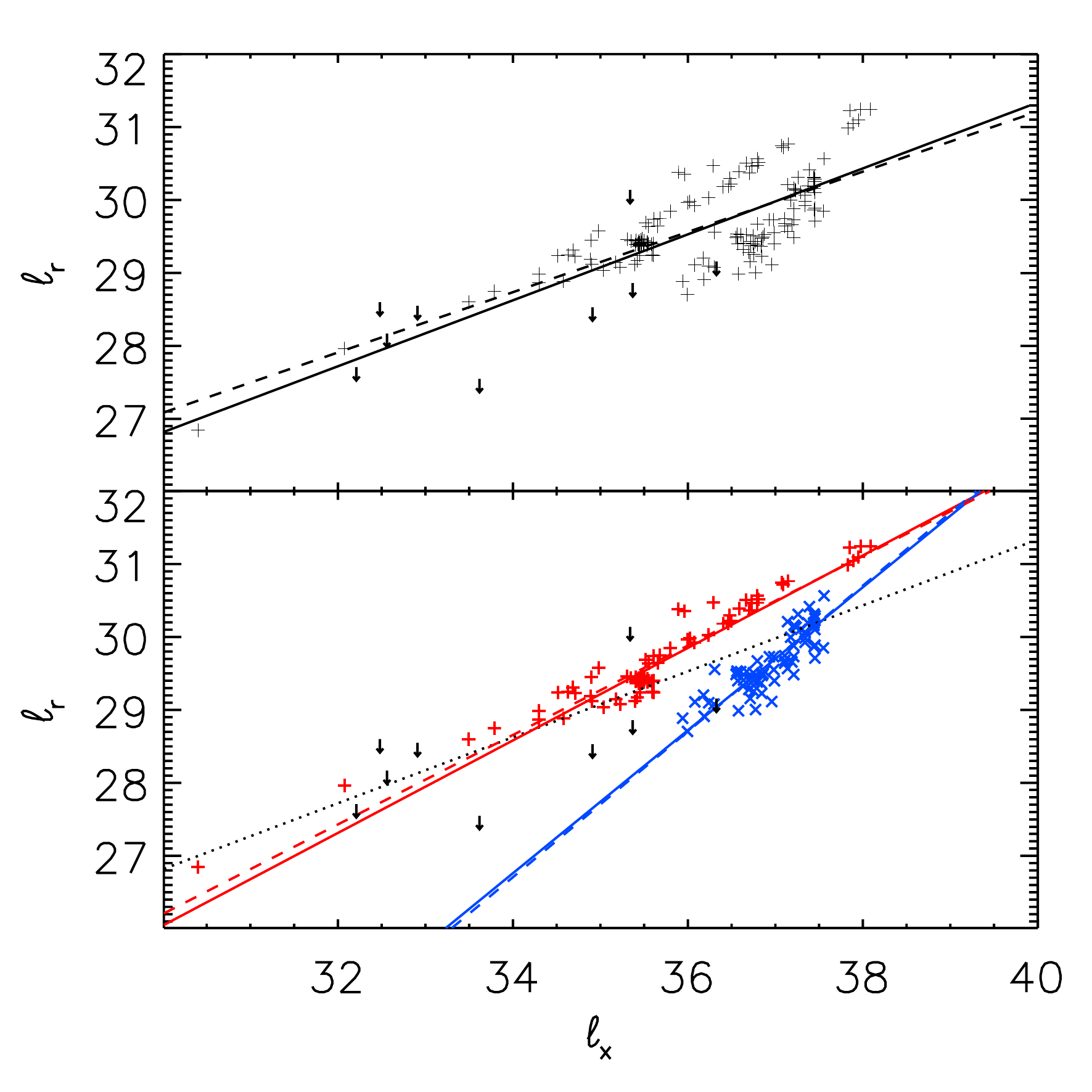,width=0.475\textwidth,angle=0}
\caption{\small Results from the linear regression analysis, performed with the Bayesian modeling package of Kelly (2007). The solid
(dashed) lines are computed for all data points (detections
only). Top: full dataset. Bottom: fit to two clusters. See Figure~\ref{fig:parplot} for details on the best-fit parameters.}\label{fig:fitplot}
\end{figure}

The Kelly (2007) code permits input of measurement errors and optional
dependent variable censoring. We assume uncertainties on both
${\ell}_{\rm x}$ and ${\ell}_{\rm r}$ of 0.15 dex. The radio upper limits are included
in {\it all} fits, but fitting is also carried out using only the
detected points for comparison and for ${\chi}^2$ tests. Three
Gaussians are used in the independent variable mixture modeling, and a
minimum of 5,000 iterations are performed with Gibbs sampling. The most
likely parameter values are estimated as the {\it median} of 10,000 draws
from the posterior distribution, with credible intervals corresponding
to 1$\sigma$ errors calculated as the 16th and 84th percentiles. The
best-fit models for each cluster are calculated separately.\\
The Hall (2011) code provides a framework
for using adaptive Markov Chain Monte Carlo to optimize parameters for a user-specified model and priors. It constructs a linear model for which the prior
probabilities of $\alpha$ and $\beta$ within each considered cluster
are taken to be Gaussian with mean zero. After each set of iterations the program conducts
self-diagnostics and if necessary suggests an additional run with
modified sampling techniques to ensure convergence. The final runs
consisted of 40,000--100,000 iterations, thinned to retain 1,000
points. The most likely parameter values are again estimated as the
medians. For the purpose of this work, the Hall code is employed in order to compare the results from the Kelly code {with detections only} (this is because the Hall code would require modifications in order to deal with upper limits and simultaneously evaluate the goodness of fit).

\begin{figure}
\psfig{figure=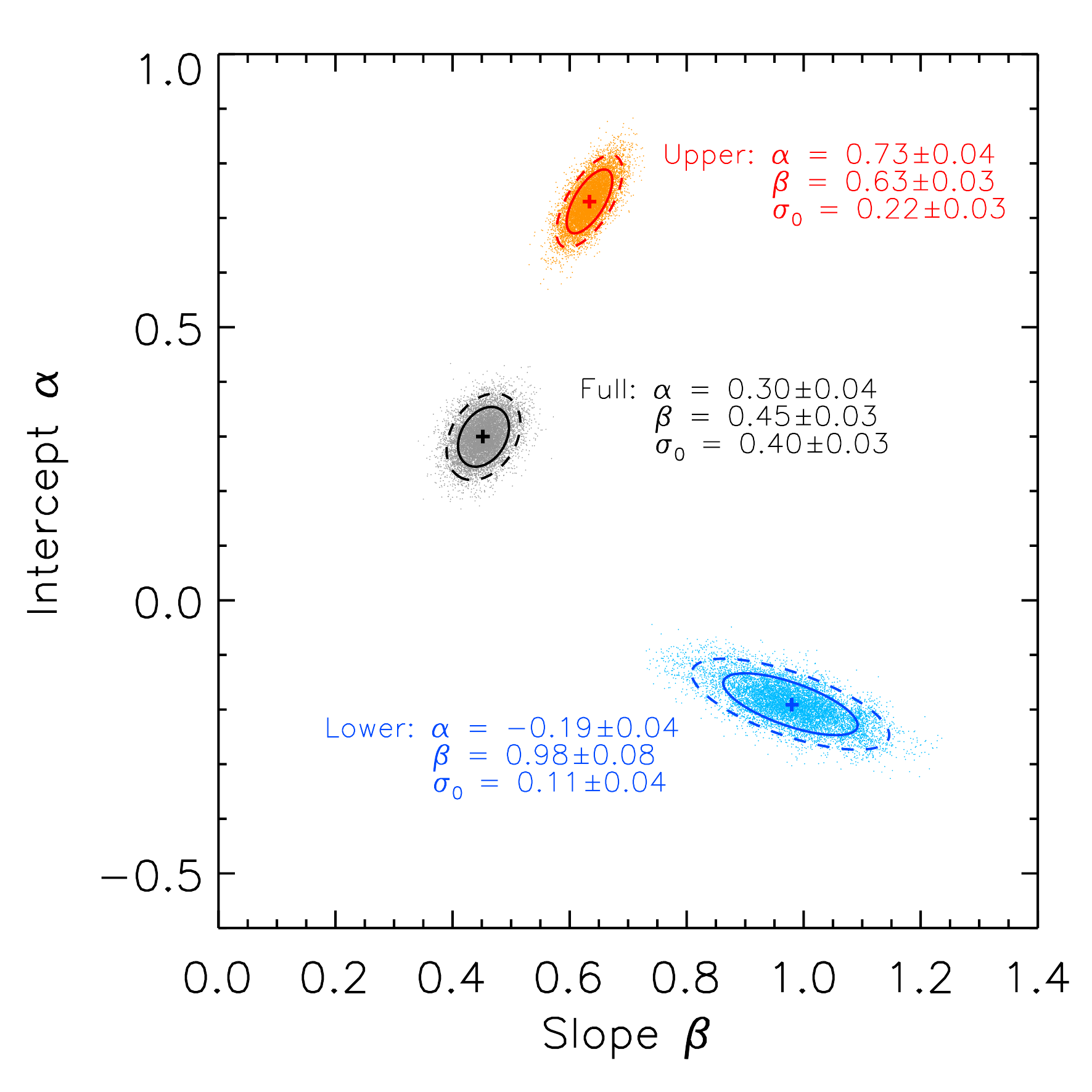,width=0.475\textwidth,angle=0}
\caption{\small Results from the linear regression analysis conducted in Section 4 (using
the code by Kelly 2007). The ellipses enclose 68 per cent  (solid) and 90 per cent 
(dashed) joint confidence regions for the intercept ($\alpha$) and the
slope ($\beta$). The crosses are the best-fit values. Quoted errors on the
fitted intercept, slope, and intrinsic scatter (${\sigma}_{\rm 0}$) are at
the 1$\sigma$ level for one parameter of interest. Label colors correspond
to the fitted models in Figure~6, with black for the one cluster model and
red/blue for the upper/lower track in the two cluster model. The intrinsic
scatter is lower for the two cluster model, with ${\sigma}_{\rm
0}=0.23/0.10$ versus ${\sigma}_{\rm 0}=0.40$ for the one cluster model. 
}\label{fig:parplot}
\end{figure}

We consider three models: one cluster, two clusters, or three clusters -- the latter as identified by {\it hybridHclust} and {\it apcluster}. 
The various models are compared through both ${\chi}^2$
versus degrees-of-freedom tests (implemented in IDL by Craig
Markwardt as {\it mpftest.pro\/}), and through Laplace-Metropolis
estimation of Bayes factors, for the Kally and Hall code, respectively.
Both $f$-test (Kelly) and Bayes factor (Hall) calculations indicate that
{\it the two cluster model, with independent linear fits, is a
significant improvement over fitting all points as a single cluster}
($>$$99.9$\% confidence). This is reflected in the highly reduced values of the intrinsic scatter, $\sigma_0$, which decreases from $0.40$ for the one cluster model to $0.22/0.11$, respectively, for the higher and lower track fits for the two cluster model. Results of the linear regression are shown in Figures~\ref{fig:fitplot} and~\ref{fig:parplot}. 
The slope of the linear fit to the upper track in the two
cluster model (from Kelly's code, including upper limits) is $0.63\pm0.03$, while the slope for
the lower track is steeper, with $0.98\pm 0.08$. 
The Hall code gives similar slopes, with 0.60$\pm$0.04 and 0.88$\pm$0.13, respectively for the upper and lower track for {\it detected} points. 
The Kelly code returns consistent values with detections only: the fitted slopes are 0.61$\pm$0.02 and 0.99$\pm$0.08 (showing that the upper limits are almost completely uninformative for the lower track).  This indicates that the few upper limits do not affect the results, providing support for the robustness of the clustering analysis from which they were necessarily excluded.  However, we wish to stress that the uncertainties on the linear regression parameters are likely underestimated in that they are {\it conditional on the cluster identifications}. 

As an additional sanity check, we randomly selected
two subsamples, with sizes corresponding to those of the two genuine
clusters, and verified that in this case the dual line fit was
properly rejected by both methods as a potential significantly
improved model. The $f$-test calculations further suggest that the
three cluster model, with independent linear fits, is at most a
marginal improvement ($p=0.04$) over the two cluster model and fits, 
although the Bayes factor suggests a more significant
improvement. Regardless, the potential improvement for the three
cluster model is achieved through a flattening of the low-luminosity
segment of the upper track, which may be influenced by the location of
the many XTE J1118+480 points slightly below the upper
track. Qualitatively similar results hold for the subset of objects
with secure distance measurements, and in particular a separate line
through the six lower track points provides a significant improvement
to the overall fit.

\section{Summary and Discussion}

Since the claim of a universal correlation between the radio and X-ray luminosity for hard state BHBs (GFP03), a more complex pictures has emerged. 
With the addition of newly discovered sources, as well as new data from well studied ones, the scatter about the 2003 best-fit relation has been steadily increasing. 
In fact, the analysis presented here shows that, from a statistical point of view, {\it the existence of a single universal
correlation between the radio and X-ray luminosity of hard and quiescent state BHBs
is to be dismissed}. Our analysis, based on a number of clustering algorithms applied to nearly-simultaneous radio/X-ray observations of 18 BHBs, shows that a two cluster model, with independent linear fits, provides a better
description of the data. The existence of two tracks does not depend on whether
sources with highly uncertain distances are excluded from the
analysis, although the actual functional form of the fits does, particularly for the lower track. 
The slope of the linear fit to the upper track ($0.63\pm0.03$) is still consistent, within the errors, with the best-fit slope ($0.7\pm 0.1$) given by GFP03 (later revised to a shallower $\simeq 0.6$ by Gallo \etal 2006 and Corbel \etal 2008), as well as with the fitted slope  for the black hole Fundamental Plane (FP) relation ($0.60\pm0.11$, although a small fraction of the BHB data considered here is generally included in the whole FP sample). The lower track slope ($0.98\pm 0.08$) is not consistent with the upper track, nor it is with the widely adopted value of $1.4$ for the neutron stars (Migliari \& Fender 2006).  

These results are qualitatively insensitive to selecting sub-samples, to the treatment of upper limits, and, we argue below, to accuracy in the source distances. 
Uncertain distances naturally tend to be equal to or larger than Galactic Center distance, where absorption is more severe (this is true for 6 out of the 7 sources in the uncertain distance sample, i.e. with the exception of XTE J1650$-$500\footnote{The generally adopted distance of 2.6 kpc is simply based on the assumption that it reached a certain fraction of the Eddington luminosity during a soft-to-hard state transition (Maccarone 2003).}). Notwithstanding this effect, the lower track is very unlikely to be purely distance-driven. If distances were systematically over-estimated for the uncertain distance sample, this would make them even less luminous, effectively widening the divide with the upper track. On the other hand, as a result of the {linearity} in the lower track slope, inflating the uncertain distances of the lower track sample to larger yet values would shift the lower track towards higher luminosities {\it along} the  $L_{\rm X}\propto L_{\rm r}$ line. The average distance of the uncertain distance sample would have to be increased to $\simgt 20$ kpc in order to have the lower track data points merge with the most luminous data points within the higher track, which seems unlikely, at least for {\it all} the lower track sources. 

In our analysis, we purposely chose not to group data points belonging
to the same source, such that the clustering analysis is somewhat
blind to specific sources. An orthogonal approach is that followed by Soleri \& Fender (2011), who, however, could not 
identify any obvious dependence of the BHB jet power (as traced by
their radio luminosity) on the binary system parameters (i.e. orbital
period, disc size, inclination) and/or outburst properties.  Although Soleri \& Fender are able to
reproduce the large scatter observed in the radio/X-ray luminosity
plane by inflating the jet bulk Lorentz factors to values larger unity for X-ray luminosities 
above 10 per cent of the Eddington limit and assuming
random inclinations, this scenario is admittedly contrived and even at
odds with some of the measured inclination angles. 

Rather than a single correlation with large scatter, the analysis
presented here provides strong evidence for the existence of two
tracks in the radio/X-ray luminosity plane. The actual cluster composition differs somewhat 
from that suggested by Coriat \etal (2011) based on visual
inspection of the radio/X-ray luminosity plane for a sub-sample of
BHBs plus the neutron stars, particularly when it comes to the top right corner of the plane (specifically, Coriat \etal single out all of the H1743$-$322 high-luminosity points as belonging to their lower track -- see figure 5 in their paper; see also Kalemci \etal 2006 and Jonker \etal 2010). Most importantly, the appealing suggestion that the lower track may be related to radiatively efficient accretion in the hard state strongly relies on a slope of $\sim 1.4$ for the lower track\footnote{The suggestion that the lower track might be related to radiatively efficient accretion (Coriat \etal 2011) comes from the combination of: (i) the theoretical scaling between the radio and the total jet power: $P_{\rm r}\propto L_{\rm jet}^{1.4}$ for partially self-absorbed jets (Blandford \& K\"onigl 1979; Falcke \& Biermann 1996; Heinz \& Sunyaev 2003); (ii) the empirical relation $L_{\rm r} \propto L_{\rm X}^{1.4}$. Under the assumption that the total jet power is directly proportional to the mass accretion rate, $\dot{m}$, the two scaling relations combine to give $L_{\rm X}\propto {\dot{m}}$ for the lower track, characteristic of radiatively efficient accretion. Similarly,  $L_{\rm r} \propto L_{\rm X}^{0.7}$ leads to $L_{\rm X}\propto {\dot{m}}^2$ for the upper track, which would be indicative of radiatively inefficient accretion.}, which is not consistent with the value derived in this work based on a larger sample and more rigorous analysis.

Although there is no reason to believe that the two luminosity tracks 
may be a selection effect due to lack of observational coverage in the
gap, it should be noted the lower track occupies a region of the parameter space at relatively high X-ray luminosities, a fact which led Coriat \etal (2011) to suggest that the two tracks may still collapse back to one at Eddington-scaled X-ray luminosities below $\sim10^{-3.5}$.  While this seems to be the case based on the behavior of H1743$-$322, which appears to have made a transition from the lower to the higher track as its X-ray luminosity decreased,  there is a small but not negligible chance that the low X-ray luminosity, upper track data points for this source may actually correspond to unresolved optically thin radio ejections (see discussion in Jonker \etal 2010), rather than to partially self-absorbed core radio emission from a steady jet (if so, they ought to be discarded). Unfortunately, as discussed by Miller-Jones \etal (2011), the current upper limits on the radio luminosities of most BHBs (see Figure 1) make it impossible to establish whether the divide may also extend toward very low Eddington ratios, although the low upper limits on the radio luminosity of GRO~J1655$-$40 and (particularly) XTE J1550$-$564 in quiescence reported by Calvelo \etal (2010) indicate that that might be the case. 

On the other hand, there are at least {\it two} sources that seem to be able to make transitions
between the two tracks:  beside H1743$-$322 (with the aforementioned caveats), GRO~J1655$-$40, too, has two simultaneous radio-X-ray detections each falling on either track (and so does the newly discovered XTE J1752$-$522; Ratti et. al., submitted). 
While better sampling is clearly needed in order to make a stronger statement than based on a handful of data points, we conclude that there is {\it tentative evidence for sources being able to make transitions from the lower to the upper track as they fade towards quiescence}, although with no obvious parameter responsible for driving such a transition. 
While the option remains open that {\it some} of the
lower track sources may actually be neutron stars, XTE~J1550$-$564, XTE~J1650$-$500 and GRO~J1655$-$40 (only one out of two data points) all sit on the lower track, and all have confirmed black hole primaries~(Orosz \etal 2011; Orosz \etal 2004; Greene \etal 2001), ruling out the possibility that {\it all} the sources on the lower track may be neutron stars. If confirmed, sources `jumping' between the two tracks would further validate this statement. 

We have no compelling explanation for this behavior, not for the very existence of the two luminosity tracks. Hysteresis effects, of the same kind as reported for the near-IR-X-ray correlation by Russell \etal (2007), are not responsible for the observed pattern. As an example, GX339$-$4 is known to fall nicely on the upper track over repeated outbursts cycles, both during the rise and the decline (e.g. Corbel \etal 2003). 
Casella \& Pe'er (2009) ascribe the large scatter about the correlation to differences in the strength of the jet magnetic field from source to source. However, there is no straightforward explanation for a dual track behavior in the context of this model (Pe'er \& Casella 2009), nor for the jumping behavior. Similarly, the existence of two tracks with different slopes is not expected within the jet-accretion models of Markoff \etal (2001,2003) and/or Yuan \etal (2005). The latter predicts that the slope of radio:X-ray correlation should become steeper ($\simeq$$1.2$), though at Eddington-scaled X-ray luminosities below $\simlt 10^{-5}-10^{-6}$, while we observe dual tracks at  Eddington-scaled X-ray luminosities above $\sim$$10^{-4}$ (in this respect, deeper observations of quiescent BHBs are surely desirable in order to establish whether the lower track extends down to very low luminosities, as results by Calvelo \etal seem to indicate). 

Finally, as also shown by Fender \etal (2010), the existence of two tracks within the radio/X-ray luminosity domain of BHBs cannot be possibly driven by two BHB sub-samples with low and high spin. This conclusion remains valid even restricting the analysis to spin measurements from reflection-fitting methods alone (e.g. 4U~1543$-47$ is the most radio loud source in our sample -- in terms of radio to X-ray luminosity ratio -- and has an inferred spin of $0.3\pm0.1$; the opposite is true for GRO~J1655$-$40, on the lower track with an estimated spin value of $0.98\pm0.01$; Miller \etal 2009) or to thermal continuum-fitting alone (e.g. A0620$-$00 lies on the low luminosity end of the upper track, with a spin value of $0.12\pm0.2$ while GRO~J1655$-$40 is on the lower track with a spin value between 0.65 and 0.75; Gou \etal 2010, Shafee \etal 2006, respectively). Furthermore, even in the absence of $any$ spin measurement, sources jumping between tracks rules out the possibility the black hole spin has a strong effect on the radio luminosity of the compact jet  and hence on defining the two tracks, since the spin value of any given source can not possibly vary on the typical time scales of BHB outbursts (here we strictly refer to radio emission from the partially self-absorbed core radio jet observed in the hard state, a.k.a. compact jet -- while it should be noted that a different conclusion has been drawn by Narayan \& McClintock 2012 based on the luminosity peak of radio flares associated with hard-to-thermal state transitions). 

Our results confirm and strengthen the conclusions of Soleri \& Fender (2011) that the existence of two separate tracks in the radio X-ray radio domain of BHBs does not reflect any obvious trend in any of the (known) physical parameters of the systems.  
The possibility remains open that the two tracks may reflect substantial differences in the actual average X-ray energy spectra. Pending availability of pointed X-ray observations with simultaneous radio coverage, a detailed X-ray spectral study could potentially highlight systematic differences between the two tracks, e.g. in terms of curvature in the hard X-ray spectrum and/or presence/absence of a optically thick disk (see Miller \etal 2006; Tomsick \etal 2009; Cabanac \etal 2009; Dunn \etal 2011).

While a theoretical interpretation of this results is beyond
the scope of this work, the existence of two tracks in the radio/X-ray luminosity domain of BHBs
significantly hampers the predictive power of any X-ray binary
radio/X-ray flux measurement for the purpose of mass and/or distance
determination. Also, it should be kept in
mind that these two tracks are not to be considered as analogs to the
radio-quiet and radio-loud tracks (if any) for AGN: as demonstrated by K\"ording \etal (2006b) the entirety of the
observations presented here -- i.e. those of hard state BHBs --
closely resemble AGN of the `radio-quiet' variety, i.e. have moderate Eddington
scaled X-ray luminosities 
and fairly stable jets, with partially self-absorbed spectra
reminiscent of compact extra-galactic radio cores. By contrast,
radio-loud AGN would be comparable to BHBs in the very high state,
showing powerful and transient jet ejections with optically thin
spectra, while BHs accreting between a few and 10 per cent of the Eddington X-ray luminosity are known to display a very different pattern radio:X-ray behavior (e.g. Chatterjee \etal 2009, King \etal 2011, and references therein).  

Further extrapolating these results to AGN and the FP of black hole activity, we argue that, while the
measured scatter to the plane potentially allows for multiple
tracks having comparable separations as the BHB's, dealing with a third
parameter (and its large uncertainties) introduces a further level of
complexity in the problem. Good prospects for breaking the mass degeneracy come from
selecting a sub-sample of super-massive black holes with dynamically
measured masses (G\"ultekin \etal 2009), since dramatically reducing the
errors on the mass estimate results in a much tighter and better
calibrated relation. An additional complication is introduced by the possibility that AGN obey the FP relation only in a time-averaged fashion (Miller \etal 2010), and that monitoring super-massive black holes on time scales of months/years may introduce further deviations from the quoted best-fit line, albeit not necessarily of the same nature as the dual tracks discussed in this work for BHBs.

\section{Acknowledgements}

B.P.M. is supported by NASA through Chandra Award Number 11620915
issued by the Chandra X-ray Observatory Center. We thank David
Childers and the UM Center for Statistical Consultation and Research
for useful suggestions.

\end{document}